# Evolution of In-Plane Magnetic Anisotropy In Sputtered FeTaN/TaN/FeTaN Sandwich Films

H. B. Nie [1], C. K. Ong [1], J. P. Wang [2] and Z. W. Li [3]

[1] *Centre for Superconducting and Magnetic Materials, Institute of Engineering Science and Department of Physics, National University of Singapore, 2 Science Drive 3, Singapore 117542.*

[2] *Media & Materials, Data Storage Institute, DSI Building, 5 Engineering Drive 1, Singapore 117608*

[3] Temasek *Laboratories, National University of Singapore, 10 Kent Ridge Crescent, Singapore 119260.*

FeTaN/TaN/FeTaN sandwich films, FeTaN/TaN and TaN/FeTaN bilayers were synthesized by using RF magnetron sputtering. The magnetic properties, crystalline structures, microstructures and surface morphologies of the as-deposited samples were characterized using angle-resolved M-H loop tracer, VSM, XRD, TEM, AES and AFM. An evolution of the in-plane anisotropy was observed with the changing thickness of the nonmagnetic TaN interlayer in the FeTaN/TaN/FeTaN sandwiches, such as the easy-hard axis switching and the appearing of biaxial anisotropy. It is ascribed to three possible mechanisms, which are interlayer magnetic coupling, stress, and interface roughness, respectively. Interlayer coupling and stress anisotropies may be the major reasons to cause the easy-hard axis switching in the sandwiches. Whereas, magnetostatic and interface anisotropies may be the major reasons to cause biaxial anisotropy in the sandwiches, in which magnetostatic anisotropy is the dominant one.

E-mail: hongbinnie@yahoo.com; *scip8379@nus.edu.sg*. Fax: *65-67776126*.

As promising thin film write head candidates for ultra-high density/data-rate magnetic recording, iron-based nitride multilayers [1-4] and sandwich structures [5-7] have attracted much attention recently. In our previous work [8] we have observed that, the direction of easy axis switches 90º when the film is thick enough in single-layered FeTaN films, and have seen an evolution of in-plane magnetic anisotropy in [FeTaN/TaN]$_n$ multilayers. For a better understanding of the underlying mechanism of the evolution of in-plane magnetic anisotropy, we examine sandwich layers existing of two ferromagnetic layers separated by one non-magnetic interlayer only, instead of larger multilayer structures. In the present paper, we report our results on the evolution of the in-plane anisotropy in FeTaN/TaN/FeTaN sandwich films. To the authors' knowledge, this phenomenon has not been observed before. Our work may shed light on the mechanism of in-plane anisotropy evolution in these sandwiches.

FeTaN/TaN/FeTaN sandwich films, FeTaN/TaN bilayers (TaN is capping layer) and TaN/FeTaN bilayers (TaN is buffer layer) were synthesized by using reactive RF magnetron sputtering on Si substrates. The description of the experiments may be found in our previous work [8, 9]. In the series of thin films, the thickness of FeTaN layers was kept at 30±3 nm, while the thickness of TaN layers, $t_{TaN}$, was varied from 0.0 nm to 50 nm. We characterized the as-deposited samples using an angle-resolved M-H loop tracer, vibrating-sample magnetometry (VSM), X-ray diffraction (XRD), transmission electron microscopy (TEM), auger electron spectroscopy (AES) and atomic force microscopy (AFM) to study their magnetic properties, crystalline structures, microstructures and surface morphologies. To reveal the magnetic anisotropy of the films, we studied the evolution of *M-H* hysteresis loops as a function of the applied field angle ϕ using the M-H loop tracer under the maximum field of 50 Oe. The angle ϕ is the angle between the directions of the applied magnetic field $\vec{H}_m$ for the M-H loop measurement and the aligning magnetic field $\vec{H}_{al}$ applied during film deposition. Each sample was measured under varied angle ϕ from 0° to 360°, at 15° intervals. Out-of-plane magnetic hysteresis loops were measured by VSM under the maximum field of 14000 Oe.

The AES depth profile of the samples indicates the sandwich structures of the films. After comparing the in-plane and out-of-plane M-H loops of the samples we conclude that there is no perpendicular magnetic anisotropy in all of the films, which is expected because the demagnetizing fields of the thin-film geometry effectively force magnetization to lie in the film plane. However, the angle dependence of in-plane coercivity of the samples showed an evolution of in-plane anisotropy for the FeTaN/TaN/FeTaN sandwich films. Figure 1 shows the polar plots of coercivities as a function of the applied field angle ϕ for the sandwiches. When $t_{TaN}$ = 0.0 and 1.0 nm, there was uniaxial anisotropy along the direction of $\vec{H}_{al}$ (ϕ = 0°). This is due to the magnetic field induced anisotropy. When $t_{TaN}$ = 2.0 nm, the easy axis switched 90° and was along the direction of ϕ = 90°. When $t_{TaN}$ = 3.0, 4.0 and 5.0 nm, the easy axis switched 90° back again and was along the direction of ϕ = 0°. When $t_{TaN}$ = 6.0 nm, the in-plane anisotropy almost disappeared, it was close to isotropy. When $t_{TaN}$ = 7.0 nm, biaxial anisotropy appeared, there were two easy axes along the direction of ϕ = 0° and ϕ = 90° respectively. When $t_{TaN}$ = 10, 20, 30, 40 and 50 nm, the films had biaxial anisotropy also. We suggest that the evolution of in-plane anisotropy is caused by the combination of three effects, which are interlayer magnetic coupling, stress and interface roughness, respectively.

Figure 2 shows the hysteresis loops of the FeTaN/TaN/FeTaN sandwich films at the angle of ϕ = 0° and ϕ = 90°, respectively, where $t_{TaN}$ changed from 0.0 nm to 50 nm. When $t_{TaN}$ = 0.0 and 1.0 nm, there were no steps in the two hysteresis loops *MH*(ϕ=0°) and *MH*(ϕ=90°) when applied field H was close to the coercivity $H_c$. In this case, there is only ferromagnetic coupling between the two FeTaN-layers. When $t_{TaN}$ was between 2.0 nm and 7.0 nm, there were no steps in the hysteresis loops *MH*(ϕ=0°), but there was a step in the hysteresis loops *MH*(ϕ=90°). The sharpest step was in the case of $t_{TaN}$ = 3.0 nm. This behavior of the magnetization is ascribed to a combination of antiferromagnetic and magnetostatic coupling between the two FeTaN-layers. Due to the partial antiferromagnetic coupling between the two FeTaN-layers, the two FeTaN layers' magnetizations do not reverse simultaneously during the reversed magnetic field process at ϕ=90°. This would occur as one of the layers abruptly reverses as the field is decreased due to this coupling, whereas it takes more energy to reverse the second layer. When $t_{TaN}$ = 3.0 nm, there is the strongest antiferromagnetic coupling between the two FeTaN-layers. When $t_{TaN}$ was between 10 and 50 nm, there were no steps in the two hysteresis loops *MH*(ϕ=0°) and *MH*(ϕ=90°) and the two loops were almost identical. This indicates that there is only magnetostatic coupling between the two FeTaN-layers with the increasing of the spacer-layer thickness. The changing of interlayer magnetic coupling with $t_{TaN}$ is consistent with the changing of coercivities with $t_{TaN}$ (see fig. 3a). Figure 3a shows two coercivities $H_c(\phi=0^o)$ and $H_c(\phi=90^o)$ as a function of $t_{TaN}$. Both the two coercivities $H_c(\phi=0^o)$ and $H_c(\phi=90^o)$ decreased with $t_{TaN}$ increasing. The decreasing coercivities were due to the reduction of domain-wall energy caused by the magnetostatic coupling between the two FeTaN-layers through interlayer TaN [1, 10-13]. Interlayer magnetic coupling will cause the evolution of the in-plane anisotropy in the sandwiches. We call such anisotropy an interlayer coupling anisotropy, which includes exchange anisotropy and magnetostatic anisotropy.

As revealed by XRD, the dominating crystalline component of the FeTaN layer in the sandwiches is nano-sized crystalline grains of *bcc* α-Fe, which show only (110) peaks in the XRD patterns. We derived the interplanar spacing of the (110) planes, $d_{(110)}$, from the XRD data. With $t_{TaN}$ increasing from 0.0 to 50 nm, the corresponding $d_{(110)}$ slightly decreased (Fig. 3b). A linear dependence between lattice strain and film stress should hold. The change of the $d_{(110)}$ may yield the change of the stress distribution in the films, therefore resulting in a change of stress anisotropy in the sandwich films. This effect in turn could cause the evolution of the in-plane anisotropy. Stress anisotropy can cause easy-hard axis switching in the thin films [14]. Interlayer coupling and stress anisotropies may be the major reasons to cause the easy-hard axis switching in the sandwiches.

To study the interface roughness effect on the in-plane anisotropy of the sandwiches, the surface morphology of the FeTaN/TaN bilayers was measured by AFM. The root mean square roughness Rq of the bilayers decreased a little with the $t_{TaN}$ increasing (Fig. 3c). The roughness varies from 0.46 nm to 0.32 nm, or a grant total of 0.14 nm. This changing is small. This is confirmed by the low magnification TEM cross-sectional images of [FeTaN/TaN]$_5$ multilayers [15]. The images reveal uniform thicknesses for each single layer and very flat interfaces between FeTaN and TaN layers. Subsequently, $H_c(\phi=0^o)$ and $H_c(\phi=90^o)$ do not vary significantly by changing $t_{TaN}$ for both kind of FeTaN/TaN and TaN/FeTaN bilayers. However, each type of FeTaN/TaN and TaN/FeTaN bilayers has different shapes of $H_c(\phi)$ polar plots although both do not vary with $t_{TaN}$. Notice that the $H_c(\phi)$ polar plots for FeTaN/TaN bilayers are all horizontal ellipse-like (the horizontal axis is larger than the vertical axis) while TaN/FeTaN bilayers are all horizontal dumbbell-like. Similarly, the strength of in-plane uniaxial anisotropy along the direction of $\bar{H}_{al}$ ($\phi = 0^o$) is almost constant as well with different $t_{TaN}$ for each type of bilayers. Usually, the TaN/FeTaN bilayers have stronger uniaxial anisotropy than that of the FeTaN/TaN bilayers. According to the above information, we may infer that interface anisotropy changed insignificantly by varying $t_{TaN}$ and its contribution to the evolution of the in-plane anisotropy in the sandwiches is sufficiently small. On the other hand, when $t_{TaN}$ is between 10 and 50 nm, there is only magnetostatic coupling between the two FeTaN-layers, and the magnetostatic coupling will cause magnetostatic anisotropy in the sandwiches. We suggest that magnetostatic anisotropy may be the dominant reason to cause biaxial anisotropy in the sandwiches.

In conclusion, the evolution of the in-plane anisotropy with the changing of $t_{TaN}$ in the FeTaN/TaN/FeTaN sandwiches, such as the easy-hard axis switching and the appearing of biaxial anisotropy, was observed. It is ascribed to three possible mechanisms, which are interlayer magnetic coupling, stress and interface roughness, respectively. Interlayer coupling and stress anisotropies may be the major reasons to cause the easy-hard axis switching in the sandwiches. Whereas, magnetostatic and interface anisotropies may be the major reasons to cause biaxial anisotropy in the sandwiches, in which magnetostatic anisotropy is the dominant one.

**References**

[1] M. H. Kryder, S. Wang and K. Pook, J. Appl. Phys. **73**, 6212 (1993).
[2] M. Naoe and S. Nakagawa, J. Appl. Phys. **79**, 5015 (1996).
[3] S. X. Li, P. P. Freitas, M. S. Rogalski, M. Azevedo, J. B. Sousa, Z. N. Dai, J. C. Soares, N. Matsakawa and H. Sakakima, J. Appl. Phys. **81**, 4501 (1997).
[4] H. Jiang, Y. J. Chen, L. F. Chen ang Y. M. Huai, J. Appl. Phys. **91**, 6821 (2002).
[5] Y. J. Chen, S. Hossain, L. Miloslavsky, Y. Liu, C. Chien, Z. P. Shi, M. S. Miller and H. C. Tong, IEEE Trans. Magn. **36**, 3476 (2000).
[6] N. X. Sun and S. X. Wang, IEEE Trans. Magn. **36**, 2506 (2000); S. X. Wang, N. X. Sun, M. Yamaguchi and S. Yabukami, *Nature* **407**, 150 (2000).
[7] T. Nozawa, N. Nouchi and F. Morimoto, IEEE Trans. Magn. **37**, 3033 (2001).
[8] H. B. Nie, S. Y. Xu, J. Li, C. K. Ong and J. P. Wang, J. Appl. Phys. **91**, 7532 (2002).
[9] H. B. Nie, S. Y. Xu, S. J. Wang, L. P. You, Z. Yang, C. K. Ong, J. Li and T. Y. F. Liew, Appl. Phys. A **73**, 229 (2001).
[10] H. Niedoba, A. Hubert, B. Mirecki and I. B. Puchalska, J. Mag. Mag. Mat. **80**, 379 (1989).
[11] J. S. S. Whiting, M. L. Watson, A. Chambers, I. B. Puchalska, H. Niedoba, H. O. Gupta, L. J. Heyderman, J-C.S. Lévy and D. Mercier, IEEE Trans. Magn. **26**, 2350 (1990).
[12] H. Clow, Nature **194**, 1035 (1962).
[13] J. C. Slonczewski and S. Middlehoek, Appl. Phys. Lett. **6**, 139 (1965).
[14] H. Deng, J. M. Jarratt, M. K. Minor, and J. A. Barnard, J. Appl. Phys. **81**, 4510 (1997).
[15] Q. Zhan, R. Yu, L. L. He, D. X. Li, H. B. Nie and C. K. Ong, "Microstructural study on multiplayer [FeTaN/TaN]$_5$ films", submitted.

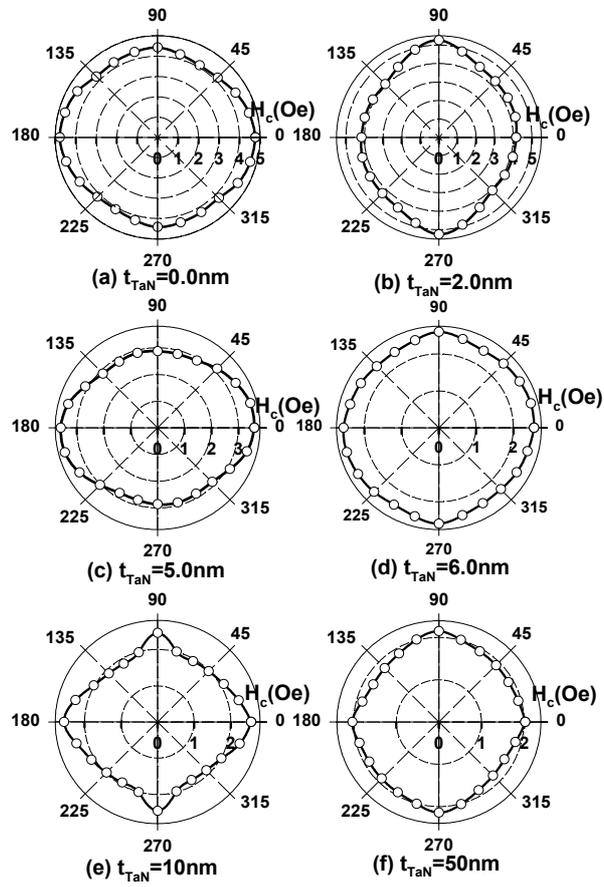

Figures 1. The polar plots of coercivities as a function of the applied field angle $\phi$ for FeTaN(30nm)/TaN($t_{TaN}$)/FeTaN(30nm) sandwich films at $t_{TaN}$ = 0.0, 2.0, 5.0, 6.0, 10 and 50 nm, respectively.

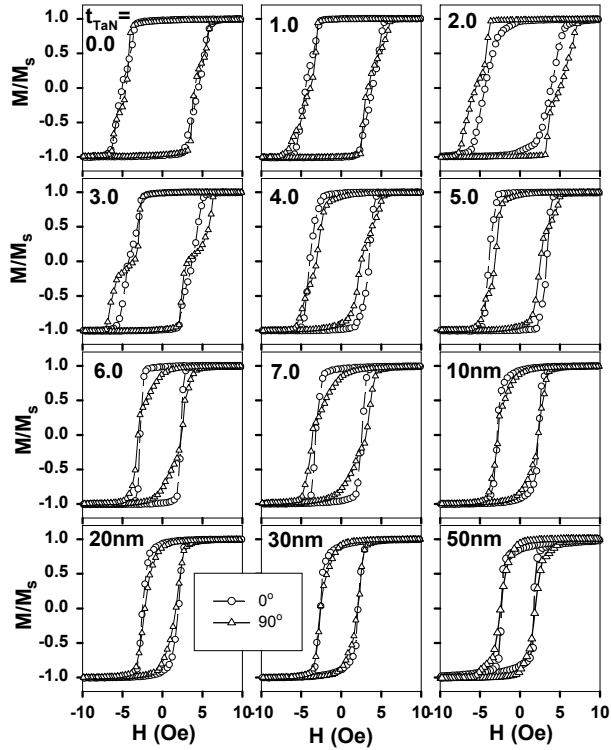

Figure 2 The hysteresis loops of FeTaN(30nm)/TaN($t_{TaN}$)/FeTaN(30nm) sandwich films at the angle of $\phi = 0°$ and $\phi = 90°$, respectively, where $t_{TaN}$ changed from 0.0 nm to 50 nm.

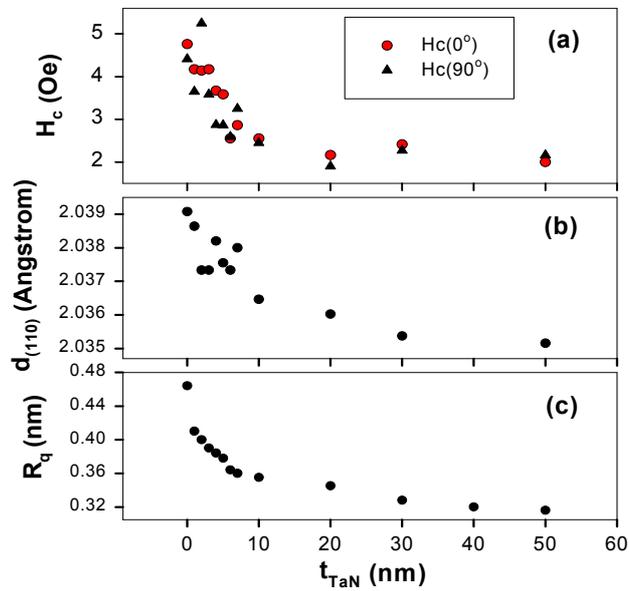

Figure 3 (a) Two coercivities $H_c(\phi=0°)$ and $H_c(\phi=90°)$ as a function of $t_{TaN}$ for FeTaN(30nm)/TaN($t_{TaN}$)/FeTaN(30nm) sandwich films. (b) The calculated lattice spacing $d_{(110)}$ of the sandwiches as a function of $t_{TaN}$. (c) The average root mean square roughness (Rq) in an area of 1.0 μm × 1.0 μm of FeTaN(30nm)/TaN($t_{TaN}$) bilayers as a function of $t_{TaN}$.